\RequirePackage{fix-cm}
\documentclass[smallextended,final]{svjour3} 
\usepackage{graphicx}
\usepackage{mathptmx}
\usepackage{url}
\usepackage{amssymb}
\usepackage{amsmath}
\def\nicefrac#1#2{\genfrac{}{}{}{1}{#1}{#2}}
\def\text#1{\mbox{\scriptsize #1}}
\def\ket#1{\mbox{$\displaystyle\vert\,#1\,\rangle$}}

\def\dyadic#1{\overset{\text{\scriptsize$\leftrightarrow$}}{#1}}
\def\vv{\mathbf{v}}
\def\vr{\mathbf{r}}
\def\vp{\mathbf{p}}
\def\vz{\mathbf{z}}
\def\vx{\mathbf{x}}
\def\vm{\mathbf{m}}
\def\vn{\mathbf{n}}
\def\vl{\mathbf{l}}

\def\vk{\mathbf{k}}

\def\vB{\mathbf{B}}

\def\vE{\mathbf{E}}
\def\He{$^{3}$He}
\def\Hea{$^{3}$He-A}

\def\cE{\mathcal{E}}

\def\be{\begin{equation}}
\def\ee{\end{equation}}
\def\ber{\begin{eqnarray}}
\def\eer{\end{eqnarray}}

\def\point#1#2{{\tt #1}_{\mbox{\footnotesize #2}}}
\def\orbital{{\tt SO(3)_{\text{L}}}}

\def\twoDspin{{\tt SO(2)_{\text{S}_{\text{z}}}}}

\def\spin{{\tt SO(3)_{\text{S}}}}
\def\gauge{{\tt U(1)_{\text{N}}}} 
\def\parity{{\tt P}} 
\def\mirror{{\tt P}_{m}} 
\def\time{{\tt T}}
\def\chiral{{\tt C}}
 
\journalname{J. Low Temp. Phys.}
\begin{document}
\title{Electron Bubbles in Superfluid $^3$He-A}
\subtitle{\sl Exploring the Quasiparticle-Ion Interaction}
\titlerunning{Electron Bubbles in $^3$He-A}
\author{Oleksii Shevtsov \and J. A. Sauls}
\institute
{
O. Shevtsov \and J. A. Sauls \at
Department of Physics and Astronomy, Northwestern University, Evanston, IL 60208 USA\\
\email{oleksii.shevtsov@northwestern.edu, sauls@northwestern.edu}
}
\date{Received: date / Accepted: date}
\maketitle
\begin{abstract}
When an electron is forced into liquid $^3$He it forms an ``electron bubble'', 
a heavy ion with radius, $R\simeq 1.5$ nm, and mass, $M\simeq 100\,m_3$, where 
$m_3$ is the mass of a $^3$He atom.
These negative ions have proven to be powerful local probes of the physical properties 
of the host quantum fluid, especially the excitation spectra of the superfluid phases.
We recently developed a theory for Bogoliubov quasiparticles scattering off electron 
bubbles embedded in a chiral superfluid that provides a detailed understanding of the spectrum of
Weyl Fermions bound to the negative ion, as well as a theory for the forces on moving electron 
bubbles in superfluid $^3$He-A (Shevtsov et al. in arXiv:1606.06240). This theory is
shown to provide quantitative agreement with measurements reported by the 
RIKEN group [Ikegami et al., Science 341:59, 2013] for the drag force and anomalous 
Hall effect of moving electron bubbles in superfluid $^3$He-A. 
In this report, we discuss the sensitivity of the forces on the moving ion to the effective
interaction between normal-state quasiparticles and the ion. 
We consider models for the quasiparticle-ion (QP-ion) interaction, including the hard-sphere
potential, constrained random-phase-shifts, and interactions with short-range repulsion 
and intermediate range attraction.
Our results show that the transverse force responsible for the anomalous Hall effect is
particularly sensitive to the structure of the QP-ion potential, and that strong short-range 
repulsion, captured by the hard-sphere potential, provides an accurate model for computing the 
forces acting on the moving electron bubble in superfluid \Hea.
\keywords{Superfluid $^3$He \and Electron bubbles \and Scattering theory \and $t$ matrix
         \and Weyl Fermions \and Broken Parity and Time-Reversal \and Chirality}
\PACS{67.30.H- \and 67.30.hm \and 67.30.hb}
\end{abstract}

\section{Introduction}\label{intro}
\vspace*{-3mm}

Superfluid $^3$He-A films are the realization of a chiral topological superfluid \cite{vol88}.
In confined geometries superfluid $^3$He-A possesses a macroscopic ground-state angular momentum, 
$L_z = (N/2)\hbar$, where $N$ is the number of $^3$He atoms in the film. The 
currents responsible for $L_z$ originate from the spectrum of Weyl Fermions confined on the 
boundary, and reflect the broken time-reversal and mirror symmetries of the chiral 
A-phase \cite{vol92,sto04,sau11,tsu12b,tsu12}.
Experimental confirmation of these broken symmetries was
demonstrated by the RIKEN group by measuring the forces on electrons moving under the 
free surface of superfluid $^3$He-A \cite{ike13,ike13b,ike15}.
Electrons submerged in superfluid $^3$He form a polaron-like state, a negative ion, 
commonly called an ``electron bubble'', reflecting its spherically symmetric ground-state 
wave function \cite{fer57,kup61}. Electron bubbles have an effective mass $M\simeq 100\,m_3$, 
where $m_3$ is the $^3$He atomic mass, and are approximately $3\,\mbox{nm}$ 
in diameter \cite{and68}. These mesoscopic objects provide a powerful local probe of the 
excitation spectrum of the quantum fluid. 
In particular, by studying the mobility of electron bubbles in $^3$He-A, Ikegami et al. have 
demonstrated the chiral nature of this superfluid \cite{ike13}. 
Skew scattering of quasiparticles by moving electron bubbles in $^3$He-A generates a transverse 
force, and thus an anomalous Hall component in the mobility tensor \cite{she16}.
An essential ingredient to the theory is the effective potential describing the interaction
between quasiparticles and ions. The potential determines the $t$ matrix for the scattering 
of normal-state quasiparticles by the ion. 
The corresponding phase shifts for normal-state QP-ion scattering are the key input
parameters to the theory for the scattering of Bogoliubov quasiparticles by the ion in the 
superfluid phase.
For temperatures above the superfluid transition, $T_c\simeq 1\,\mbox{mK}\le T\lesssim 30\,\mbox{mK}$, 
the mobility of the negative ion is independent of 
temperature \cite{aho76a,aho78}, $\mu^{\mbox{\small exp}}_{\mbox{\small N}}$ 
$\simeq 1.7\times 10^{-6}\mbox{m}^2\mbox{/Vs}$,
and determined by the normal-state QP-ion transport cross-section, 
$e/\mu_{\mbox{\small N}} = n_3\,p_f\,\sigma_{\mbox{\small N}}^{\mbox{\small tr}}$,
where $\sigma_{\mbox{\small N}}^{\mbox{\small tr}}$ is given by Eq.~(8) of 
Ref. \cite{she16}. This relation is used to constrain models for the QP-ion potential.

We discuss sensitivity of the forces on details of the QP-ion potential. 
For the electron bubble, the simplest model of a hard sphere potential provides a good
description of both the longitudinal and transverse forces on the bubble in chiral superlfuid 
$^3$He-A \cite{she16}.
For repulsive, short-range interactions the details of the QP-ion potential are shown 
to be relatively unimportant in determining the longitudinal force on the moving electron bubble 
provided the normal-state transport cross-section accounts for the normal-state mobility. 
The transverse force is shown to be more sensitive to the structure of the QP-ion potential
and corresponding phase shifts as a function of the angular momentum channel.   
QP-ion interactions with intermediate range attraction, in addition to short-range repulsion, 
lead to significant discrepancies between theory and experiment for the 
magnitude and temperature dependence of the transverse force on moving electron bubbles.
Only models with strong repulsion at a mesoscopic distance of order the size of the bubble 
provide good agreement for both the longitudinal and transverse forces.
This explains the success of the single parameter hard-core QP-ion potential in providing 
quantitative predictions for the forces in the superfluid A-phase.

\vspace*{-3mm}
\section{Stokes drag and the anomalous Hall effect of electrons moving in $^3$He-A}\label{sec_AHE}

Superfluid \Hea\ is a condensate of equal-amplitude, spin-aligned Cooper pairs,
$\nicefrac{1}{\sqrt{2}}(\ket{\rightrightarrows}+\ket{\leftleftarrows})$,
each with an orbtial wave function - or mean-field order parameter -
$\Delta(\vp) = \Delta(\hat\vm + i \hat\vn)\cdot\vp/p_f$, where $\vp$ is the relative 
momentum of the Cooper pair. Each Cooper pair has orbital angular momentum projection $\hbar$ 
along an axis $\hat\vl \equiv \hat\vm\times\hat\vn$. The ground state spontaneously breaks 
time-reversal symmetry ($\time$), parity ($\parity$), orbital ($\orbital$) and spin ($\spin$) rotation 
symmetries in addition to gauge symmetry ($\gauge$).
However, these symmetries are only partially broken. The residual symmetry of the A phase is 
$H = \twoDspin\times\point{U(1)}{N-L$_z$}\times\chiral$, where 
$\chiral=\time\times\mirror$ is \emph{chiral} symmetry defined as the product of time-reversal 
and mirror symmetry ($\mirror$) in a plane containing the chiral axis $\hat\vl$: 
$\mirror\,\hat\vm=+\hat\vm$, $\mirror\,\hat\vn=-\hat\vn$, and thus 
$\mirror\,\hat\vl=-\hat\vl$. Similarly, $\time\,(\hat\vm+i\hat\vn)=(\hat\vm-i\hat\vn)$, 
and thus $\time\,\hat\vl=-\hat\vl$, i.e. both time-reversal and mirror symmetry are broken in \Hea, 
but chiral symmetry, $\chiral=\time\times\mirror$, is preserved.  
For our purposes the other important residual symmetry of \Hea\ is rotational symmetry about the 
chiral axis \emph{modulo} a gauge transformation, i.e. the group $\point{U(1)}{N-L$_z$}$. 
Thus, observables such as the superfluid density that are described by a rank two tensor are 
constrained to be uniaxial. 
In particular the force on an electron bubble moving with velocity $\vv$ in superfluid \Hea, 
\begin{equation}\label{F_qp}
\mathbf{F}_{\mathrm{QP}} = -\dyadic{\eta}\cdot\vv
\,,
\end{equation}
is defined, in the linear response limit, by a Stokes tensor of the form \cite{she16},
\be\label{eq-Stokes_Tensor}
\eta_{ij} = \eta_{\perp}\left(\delta_{ij} - \hat\vl_i\hat\vl_j\right)
          + \eta_{\parallel}\,\hat\vl_i\hat\vl_j
	  + \eta_{\text{AH}}\epsilon_{ijk}\hat\vl_k
\,,
\ee
where all components of the Stokes tensor are real with $\eta_{\perp}$ ($\eta_{\parallel}$) defining
the drag force for motion perpendicular (parallel) to the chiral axis. The off-diagonal term, 
$\eta_{\text{AH}}$, in the Stokes tensor gives rise to a \emph{transverse} force acting on the 
ion for motion perpendicular to $\vl$. The transverse component of the force is allowed by chiral 
symmetry, but would vanish if \Hea\ were mirror symmetric \cite{she16}.

Under the action of a uniform electic field, $\vE\perp\hat\vl$, the equation of motion for an 
electron bubble in \Hea\ is 
\be\label{EOM}
M\frac{d\mathbf{v}}{dt} = e\mathbf{E} - \eta_{\perp}\mathbf{v} 
                        - \eta_{\text{AH}}\vv\times\hat\vl
\,.
\ee
The electric field accelerates the electron bubble, which is opposed by the 
Stokes drag, $-\eta_{\perp}\vv$, and the transverse force, $-\eta_{\text{AH}}\vv\times\hat\vl$. 
The latter gives rise to an \emph{anomalous} Hall effect, characterized by an effective magnetic field,
\be
\vB_{\text{eff}} = -\frac{c}{e}\eta_{\text{AH}}\,\hat\vl
\,.
\ee
The steady-state solution for the terminal velocity is given by $0 = e\vE - \dyadic{\eta}\cdot\vv$, 
which can be inverted to give,
\be\label{Mob_def}
\vv = \dyadic{\mu}\cdot\vE
\,,
\ee
where the mobility tensor is given by 
\be
\dyadic{\mu} = e\,\dyadic{\eta}^{-1}
\,,
\ee
and has the same uniaxial structure as the Stokes tensor in Eq. (\ref{eq-Stokes_Tensor}) with
$\mu_{\parallel}=e/\eta_{\parallel}$,
$\mu_{\perp}=e\,\eta_{\perp}/(\eta_{\perp}^2+\eta_{\text{AH}}^2)$, and
$\mu_{\text{AH}}=-e\,\eta{\text{AH}}/(\eta_{\perp}^2+\eta_{\text{AH}}^2)$.
For $\vE=\cE\hat\vx\perp\hat\vl\parallel\hat\vz$ the anomalous Hall angle is given by the ratio of 
the transverse and longitudinal velocities,
\be\label{eq-Hall_angle}
\tan\alpha = \frac{v_y}{v_x} = \frac{\eta_{\text{AH}}}{\eta_{\perp}}
\,.
\ee
The experimental observation of the anomalous Hall effect for electron bubbles moving in \Hea,
including the reveresal of the Hall current under $\hat\vl\rightarrow -\hat\vl$,
provided the direct signature of chirality and broken mirror symmetry in \Hea. 
The magnitude of the effect is also remarkable, corresponding to an effective magnetic field of order 
$B_{\text{eff}}\simeq 10^3 - 10^4\,\mbox{Tesla}$.

For temperatures $0 < T < T_c$ the microscopic origin of both the drag force and transverse force
on the moving electron bubble in \Hea\ is multiple scattering of thermally excited Bogoliubov 
quasiparticles by the quasiparticle-ion potential, combined with branch conversion scattering by 
the chiral order parameter of \Hea. The formulation of the scattering theory is described in detail 
in Ref. \cite{she16}, and calculations of the structure of the electron bubble embedded in
\Hea, as well as the Stokes tensor, are reported for the hard-sphere model for the QP-ion potential
with radius $R = 1.42\,\mbox{nm}$ ($k_f R = 11.27$), and shown to be in good agreement the experimental 
results for the drag and transverse forces reported by the RIKEN group \cite{ike13,ike15} for electron 
bubbles moving in \Hea.
In what follows we discuss the sensitivity of the theoretical predictions to the QP-ion potential.
We report theoretical results for the drag and transverse forces for a wide range of models for 
the QP-ion potential and compare them with the experiments, and the one-parameter hard-sphere 
potential.

\vspace*{-3mm}
\subsection{Normal-state $t$ matrix}\label{subsec_NST}
\vspace*{-3mm}

Our theoretical description for the bound-state spectrum and transport properties of an electron
embedded in \He\ starts with a model for the \emph{effective} interaction, $U(r)$, between a 
quasiparticle and the ion, which we assume to be short-ranged and isotropic.
At short-range the potential is expected to be of order $1\,\mbox{eV}$ based on the energy 
required to form the electron bubble, while the range of the potential is to be 
of order the classical estimate of the electron bubble radius, $R\sim 2\,\mbox{nm}$. Thus, 
the theory for scattering and the transport properties of the ion is in the strong scattering 
limit for a mesoscopic object and requires a calculation of the full normal-state scattering 
$t$ matrix.
An important observation is that the scattering of quasiparticles can be treated 
in the elastic limit. The heavy mass of the electron bubble,
combined with the QP-ion collision frequency, implies that recoil of the ion is negligible, 
i.e. QP-ion scattering in normal \He\ is to a good approximation 
elastic \cite{jos69,fet77}.
In fact for the electric fields employed in the RIKEN experiments the recoilless limit
can be shown to hold down to temperatures of order $T_{\text{r}}\simeq 200\,\mu\mbox{K}$.

At the atomic level the $t$ matrix takes into account multiple scattering of \He\ atoms by 
the potential representing their interaction with the ion, and is given by a solution of
the Lippmann-Schwinger equation \cite{rammer98}, 
\be\label{LS_def}
T^R = V + VG^RT^R
\,,
\ee
where $G^R$ is the causal propagator for \He\ Fermions. 

At low temperatures, $k_BT \ll E_f$, only quasiparticle excitations with momenta near the Fermi 
surface, $\vk\simeq k_f\hat\vk$, determine the properties of $^3$He liquid. The corresponding 
excitation energies satisfy $|\xi_{\vk}| \ll E_f$. 
In the low-energy limit the equation for the $t$ matrix is obtained by isolating the
quasiparticle pole term, $G^R_{\mathrm{low}}\sim(E+i0^{+} - \xi_{\vk})^{-1}$, in 
the full propagator, $G^R = G^R_{\mathrm{low}} + G^R_{\mathrm{high}}$.
The high-energy propagator renormalizes $V$ to the QP-ion effective interaction, 
$U = V + VG^R_{\mathrm{high}}U$. This is the interaction determining the scattering
of low-energy quasiparticles by the electron bubble in normal \He.
The resulting equation for the QP-ion $t$ matrix, 
$t^R_{\text{N}}(\hat\vk',\hat\vk;E)\equiv\langle\vk'|T^R|\vk\rangle$,
describing elastic scattering of quasiparticles with energy $|E|\ll E_f$ between states with initial 
$\vk = k_f\hat\vk$ and final $\vk'=k_f\hat\vk'$ momenta is
\be\label{TM_N}
t^R_{\text{N}}(\hat\vk',\hat\vk;E) = u(\hat\vk',\hat\vk) 
+\int\frac{d\Omega_{\vk''}}{4\pi}
u(\hat\vk',\hat\vk'')\,g_{\text{N}}(\hat\vk'',E)\,
t^R_{\text{N}}(\hat\vk'',\hat\vk;E),
\ee
where 
$g_{\text{N}}(\hat\vk'',E) = N_f\int d\xi_{\vk''}\,G^R_{\mathrm{low}}(\vk'',E)=-i\pi N_f$ 
is the $\xi$-integrated quasiparticle propagator, 
$N_f = m^{\ast}k_f/2\pi^2\hbar^2$ is the single-spin density of states at the Fermi 
surface, and $m^{\ast} = p_f/v_f$ is the quasiparticle effective mass.
The matrix elements of the effective potential, 
$u(\hat\vk',\hat\vk) = \langle\vk'|U|\vk\rangle$,
as well as the $t$ matrix, are evaluated on the  Fermi surface.

The effective potential is assumed to be spherically symmetric for the ground-state of the 
electron bubble \cite{kup61}. Thus, we use standard partial-wave analysis to represent the 
$t$ matrix in terms of partial-wave amplitudes and Legendre polynomials,
\ber
u(\hat{\vk}',\hat{\vk}) 
&=& 
\sum_{l=0}^{\infty}(2l+1)u_lP_l(\hat{\vk}'\cdot\hat{\vk})
\,,
\\
t^R_{\text{N}}(\hat{\vk}',\hat{\vk};E) 
&=& 
\sum_{l=0}^{\infty}(2l+1)t^R_l(E)P_l(\hat{\vk}'\cdot\hat{\vk})
\,.
\eer
Equation (\ref{TM_N}) is then solved in terms of the $t$-matrix amplitudes, 
$t^R_l(E) = u_l/(1+i\pi N_fu_l)$, which are parametrized in terms of the scattering phase
shift, $\delta_l = -\tan^{-1}\left(\pi N_f u_l\right)$, for each angular momentum channel,
\be
t^R_l(E) = -\frac{1}{\pi N_f}e^{i\delta_l}\sin\delta_l
\,.
\ee
Note that the structure of the QP-ion potential is encoded in the set of scattering phase shifts. 
The resulting $t$ matrix determines the differential cross section for QP-ion scattering,
and thus the corresponding total and transport cross-sections,
\begin{align}
&\frac{d\sigma}{d\Omega_{\vk'}} = \left(\frac{m^{\ast}}{2\pi\hbar^2}\right)^2
|t^R_{\text{N}}(\hat{\vk}',\hat{\vk};E)|^2,\\
&\sigma^{\text{N}}_{\mathrm{tot}} = 
\int\frac{d\Omega_{\vk'}}{4\pi}\frac{d\sigma}{d\Omega_{\vk'}} = 
\frac{4\pi}{k_f^2}\sum_{l=0}^{\infty}(2l+1)\sin^2\delta_l,\\
&\sigma^{\text{N}}_{\mathrm{tr}} = 
\int\frac{d\Omega_{\vk'}}{4\pi}(1-\hat{\vk}\cdot\hat{\vk}')\frac{d\sigma}{
d\Omega_{\vk'}} = 
\frac{4\pi}{k_f^2}\sum_{l=0}^{\infty}(l+1)\sin^2(\delta_{l+1} - \delta_l).
\end{align}
The transport cross-section determines the normal-state mobility, 
$\mu_{\text{N}} = e/n_3p_f\sigma^{\text{N}}_{\mathrm{tr}}$, where $p_f = \hbar k_f$ and 
$n_3 = k_f^3/3\pi^2$ is the $^3$He particle density.

\vspace*{-3mm}
\subsection{Scattering theory for the superfluid state}\label{subsec_SSS}
\vspace*{-3mm}

The structure and transport properties of electron bubbles in \He\ are modified dramatically by
the formation of a condensate of bound Cooper pairs. Spontaneous symmetry breaking - particularly
broken gauge, parity and time-reversal in \Hea\ - has a profound 
effect on the spectral properties of the electron bubble, as well as the cross-section
for Bogoliubov quasiparticles scattering off the negative ion.
Bogoliubov quasiparticles, which are coherent superpositions of normal-state particles and holes, 
undergo branch conversion (Andreev) scattering by the chiral order parameter in combination with 
scattering by the QP-ion potential. Multiple Andreev and QP-ion scattering in \Hea\ 
leads to the formation of a bound spectrum of chiral (Weyl) Fermions, which hybridize with the 
continuum of nodal quasiparticles to form low-energy resonances with spectral weight 
confined near the electron bubble \cite{she16}.
This discrete spectrum of chiral Fermions evolves into a continuous branch of chiral edge states in the 
limit $R\rightarrow\infty$, and is a finite-size realization of the spectrum of Weyl Fermions for the 
2D topological phase of \Hea\ \cite{vol92,sto04,sau11,tsu12b,tsu12}.

Branch converstion scattering by the QP-ion potential and chiral order parameter also leads to 
skew scattering, and to an anomalous Hall effect for the motion of electron bubbles in 
superfluid \Hea\ (c.f \cite{she16} and references therein).
The $t$ matrix for normal-state spin-$\nicefrac{1}{2}$ quasiparticles is expanded to a 
$4\times 4$ Nambu matrix to encode the particle-hole coherence of Bogoliubov quasiparticles, 
\emph{and} branch conversion scattering between particle-like ($dE_{\vk}/dk > 0$) and 
hole-like ($dE_{\vk}/dk < 0$) excitations by the order parameter $\Delta(\vk)$.
The scattering theory and the transport theory for the forces on moving electron bubbles 
resulting from scattering of Bogoliubov quasiparticles is described in detail in Ref. \cite{she16}.

The Lippmann-Schwinger equation for the $t$ matrix describing scattering states in superfluid
\Hea\ can be expressed in terms of the normal-state $t$ matrix (elevated to Nambu space), and
the difference between the normal- and superfluid Nambu propagators,
\be
T_{\text{S}} = T_{\text{N}} + T_{\text{N}}(G^R_{\text{S}}-G^R_{\text{N}})T_{\text{S}}
\,.
\ee
This subtraction allows us to use the normal-state $t$ matrix, which we calculate for 
a range of models for the QP-ion potential, as \emph{input} to the calculation of 
the $t$ matrix for scatteirng of Bogoliubov quasiparticles in superfluid \He. 

The basis of scattering states is obtained by solving the Bogoliubov equation with the
pair potential defined by the chiral A-phase order parameter 
$\Delta(\hat\vp)=\Delta\sigma_x(\vp_x + i\vp_x)/p_f$, where $\vp=-i\hbar\boldsymbol{\nabla}$ 
is the relative momentum operator. We denote particle-like and hole-like Bogoliubov quasiparticle 
spinors by $|\Psi_{1,\vk\sigma}(\vr)\rangle$ and $|\Psi_{2,\vk\sigma}(\vr)\rangle$, respectively.
The total rate for QP-ion scattering with momentum change, $\vk\rightarrow\vk'$, is given 
by Fermi's golden rule, $\Gamma(\vk',\vk)=(2\pi/\hbar)\,W(\vk',\vk)\delta(E_{\vk'}-E_{\vk})$, with
\begin{align}
W(\vk',\vk) 
= 
\nicefrac{1}{2}\sum_{\sigma,\sigma' = \uparrow,\downarrow}
\Bigl\lbrack
&
|\langle\Psi_{1,\vk'\sigma'}|T_S|\Psi_{1,\vk\sigma}\rangle|^2 
+ 
|\langle\Psi_{1,\vk'\sigma'}|T_S|\Psi_{2,\vk\sigma}\rangle|^2 
\notag
\\ 
+
\,
&
|\langle\Psi_{2,\vk'\sigma'}|T_S|\Psi_{1,\vk\sigma}\rangle|^2
+
|\langle\Psi_{2,\vk'\sigma'}|T_S|\Psi_{2,\vk\sigma}\rangle|^2
\Bigr\rbrack_{E_{\vk'} = E_{\vk}}
\,,
\label{eq_W}
\end{align}
where $E_{\vk}=\sqrt{\xi_{\vk}^2 + |\Delta(\hat\vk)|^2}$ is the Bogoliubov quasiparticle 
excitation energy.
A key feature of QP-ion scattering in \Hea\ is the violation of microscopic reversibility;
the rates for QP scattering by ions embedded in superfluid \Hea\ corresponding to
momentum transfers $\vk\rightarrow\vk'$ and $\vk'\rightarrow\vk$ are not equivalent, i.e.
$W(\vk',\vk) \neq W(\vk,\vk')$.
The violation of the microscopic reversibility is a consequence of broken time-reversal 
($\time$) and mirror ($\mirror$) symmetries in the \Hea\ \cite{she16}. To highlight the importance
of the violation of microscopic reversibility on QP-ion scattering we separate the  
rate into mirror symmetric ($W^+$) and anti-symmetric ($W^-$) components,
\be
W(\vk',\vk) = W^+(\vk',\vk) + W^-(\vk',\vk)
\,,\quad
W^\pm(\vk,\vk') = \pm W^\pm(\vk',\vk)
\,.
\ee
The mirror symmetric scattering rate determines the drag force on a moving electron bubble, while
the mirror anti-symmetric rate is responsible for the transverse force, and thus the anomalous
Hall effect for electron bubbles moving in \Hea. These forces are defined in terms of the components 
of the Stokes tensor \cite{she16},
\begin{align}
\eta_{ij} = n_3p_f\int_{0}^{\infty}dE\left(-2\frac{\partial f}{\partial E}\right)\sigma_{ij}(E)
\,,\quad
\forall\, i,j\in\left\{x,y,z\right\}
\,,
\end{align}
where the components of the energy-resolved transport cross section also separate into 
symmetric and anti-symmetric tensors, $\sigma_{ij}(E)=\sigma_{ij}^{(+)}(E)+\sigma_{ij}^{(-)}(E)$,
corresponding to the signatures of $W^{\pm}(\vk',\vk)$ under $\vk'\leftrightarrow\vk$, 
\begin{align}
&\sigma_{ij}^{(+)}(E) = \frac{3}{4}\int_{E \geq 
|\Delta(\hat{\vk}')|^2}\!d\Omega_{\vk'}\int_{E \geq 
|\Delta(\hat{\vk})|^2}\!\frac{d\Omega_{\vk}}{4\pi}
[(\hat{\vk}_i' - \hat{\vk}_i)(\hat{\vk}_j' - 
\hat{\vk}_j)]\,\frac{d\sigma}{d\Omega_{\vk'}},\\
&\sigma_{ij}^{(-)}(E) = \frac{3}{4}\int_{E \geq 
|\Delta(\hat{\vk}')|^2}\!d\Omega_{\vk'}\int_{E \geq 
|\Delta(\hat{\vk})|^2}\!\frac{d\Omega_{\vk}}{4\pi}
[\varepsilon_{ijk}(\hat{\vk}' \times 
\hat{\vk})_k]\,\frac{d\sigma}{d\Omega_{\vk'}}
\left[f(E)-\frac{1}{2}\right],\label{sigma_skew}\\
&\frac{d\sigma}{d\Omega_{\vk'}}(\hat{\vk}',\hat{\vk};E) 
= 
\left(\frac{m^{\ast}}{2\pi\hbar^2}\right)^2
      \frac{E}{\sqrt{E^2-|\Delta(\hat{\vk}')|^2}}
       W(\vk',\vk)
      \frac{E}{\sqrt{E^2-|\Delta(\hat{\vk})|^2}},
\end{align}
where $f(E)$ is the Fermi-Dirac distribution function. 
Note that only the mirror symmetric (anti-symmetric) component of the scattering
rate, $W^+$ ($W^-$), contributes to the energy-resolved cross-section, 
$\sigma_{ij}^{(+)}(E)$ [$\sigma_{ij}^{(-)}(E)$]. Furthermore, $\sigma^{(+)}_{ij}(E)$ 
is a diagonal tensor, and determines \emph{only} the longitudinal drag forces on the moving
ion, while $\sigma^{(-)}_{ij}(E)$ is an anti-symmetric tensor that determines the transverse 
force, and thus the anomalous Hall current.

To compute these forces we calculate the rates, $W^{\pm}(\vk',\vk)$, based on the 
formulation outlined above and in more detail in Ref. \cite{she16}. The key input to 
the calculation is the QP-ion potential, and in particular the QP-ion phases shifts that 
define the normal-state $t$ matrix. We discuss several possible models for QP-ion
scattering below.

\section{Quasiparticle-Ion Scattering - Models and Phase Shifts}\label{sec_PS}

Scattering phase shifts are the imprint of the near-field QP-ion interaction, $U(r)$, 
on far field, asymptotic, free-particle form for the scattering solutions to the 
Schr\"odinger equation; $\psi_{lm}(\vr)=R_l(r)\,Y_l^m(\theta,\phi)$, 
where $Y_l^m(\theta,\phi)$ are spherical harmonics, and the radial wave function 
satisfies \cite{messiah58}
\be\label{Rad_schrodinger}
\frac{1}{r^2}\frac{\partial}{\partial r}\left(r^2\frac{\partial R_l}{\partial r}\right) + 
\left[k^2 - \mathcal{U}(r) - \frac{l(l+1)}{r^2}\right]R_l = 0
\,,
\ee
with $k^2 = 2m^{\ast}E/\hbar^2$ and $\mathcal{U}(r) = 2m^{\ast}U(r)/\hbar^2$. 
We consider finite-range potentials such that $U(r) \approx 0$ for $r > a$, in which 
case the radial wave function for $r >a$ is a linear combination of spherical wave 
solutions,
\be
R_l(r) = A\,[\cos\delta_l\,j_l(kr) - \sin\delta_l\,n_l(kr)]
\,,
\ee
where $j_l(kr)$ and $n_l(kr)$ are spherical Bessel functions of the first and second kind, 
respectively and $A$ is a normalization constant. The phase shift, $\delta_l(k)$, for angular 
momentum channel $l$ depends on the wavenumber, $k$; in the far field, $kr\gg 1$, 
$R_l(r) \approx A\,\sin(kr - l\pi/2 + \delta_l)/kr$, i.e. a free QP solution shifted in phase
by $\delta_l$ as a result of the near field interaction with the ion.  
Matching the near and far field solutions and the first derviatives at $r=a$ provides us with a 
normalization-independent condition for the log-derivative of $R_l$ at $r = a$, 
\be\label{phase_shifts_def}
\tan\delta_l(k) = \frac{kj_l^\prime(ka) - \gamma_lj_l(ka)}{kn_l^\prime(ka) - \gamma_ln_l(ka)}
\,,\quad
\gamma_l \equiv \frac{d\ln R_l}{dr}\Bigr|_{r = a^-}
\,.
\ee
Equation (\ref{phase_shifts_def}) can be used directly to obtain the phase shifts provided the 
near field solution $R_l(r)$ for $r < a$ can be found explicitly. 
For short-range potentials for which there is not an analytic solution we use the 
\emph{variable-phase} method to calculate the phase shifts \cite{calogero67}. This method is 
based on a first-order, nonlinear differential equation for a function, $\chi_l(r)$,
\be\label{phase_shifst_vfm}
\chi_l^{\prime}(r)=-kr^2\mathcal{U}(r)\Big[\cos\chi_l(r)\,j_l(kr)-\sin\chi_l(r)\,n_l(kr)\Big]^2
\,,
\ee
The variable-phase equation is well suited for numerical calculations. The asymptotic value obtained 
from the solution to Eq.~(\ref{phase_shifst_vfm}), subject to the boundary condition, $\chi_l(0) = 0$, 
determines the phase shift for each $l$ and $k$: $\delta_l(k) = \lim_{r\rightarrow\infty}\chi_l(r)$.

\vspace*{-3mm}
\subsection{Hard-sphere model}\label{subsec_HSM}
\vspace*{-3mm}

\begin{figure}
\centering
\includegraphics[width=1.0\textwidth,keepaspectratio]{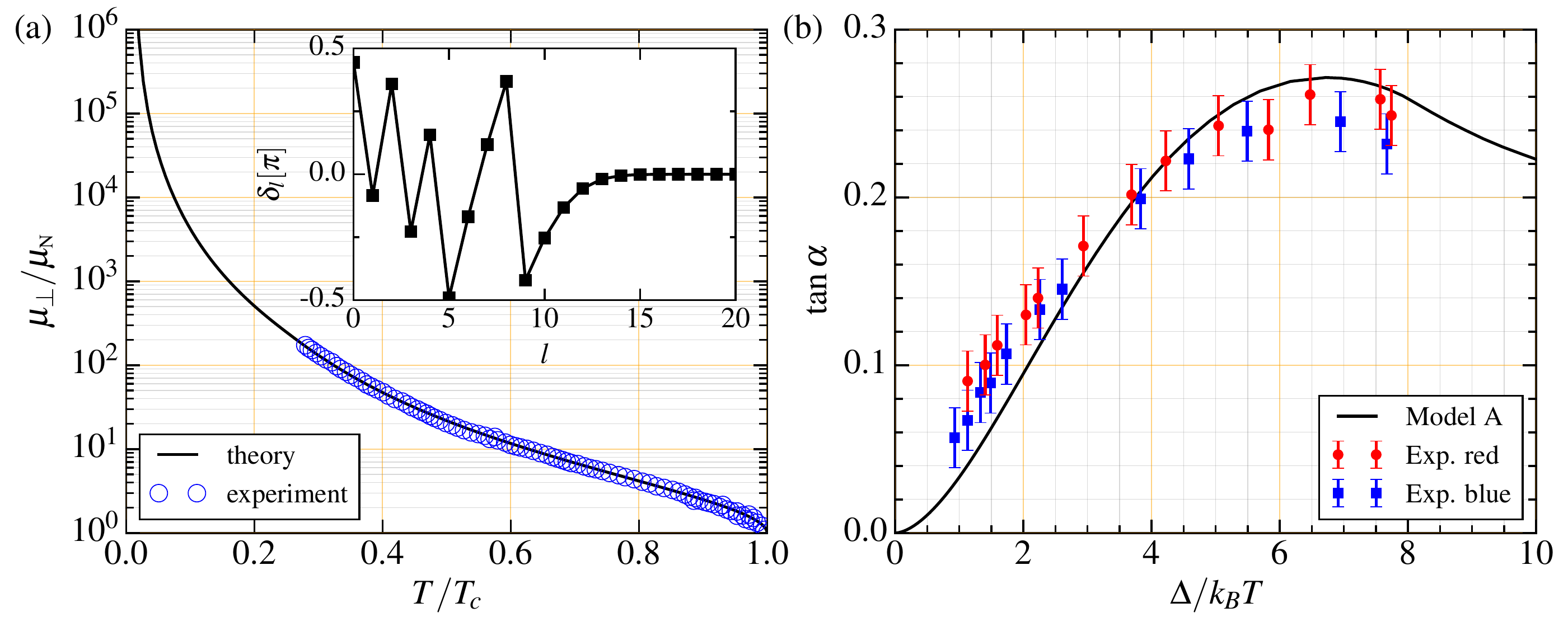}
\caption{Comparison of the hard-sphere model (Model A) with data for the mobility of 
electron bubbles in \Hea\ \cite{ike13,ike15}.
Panel (a): calculated longitudinal mobility, $\mu_{\perp}/\mu_{\text{N}}$ vs. 
$T/T_c$ (black line), with the inset showing the phase shifts for Model A. 
Experimental data shown as blue circles.
Panel (b): calculated anomalous Hall ratio, $\tan\alpha=\eta_{\text{AH}}/\eta_{\perp}$ 
vs. $\Delta(T)/k_B T$ in comparison with data from two different experimental runs 
(red and blue points) reported in Refs. \cite{ike13,ike15}.
(Color figure online.)}
\label{Fig1}
\end{figure}

The simplest model with an analytic solution for the phase shifts is the one-parameter hard-sphere 
model defined by $U(r < R) = \infty$ and $U(r > R) = 0$, where $R$ is the hard sphere radius. 
The phase shifts are found by requiring that $R_l(r=R) = 0$, and thus given by the formula, 
$\tan\delta_l(k) = j_l(kR)/n_l(kR)$ \cite{messiah58}.
The hard-sphere model provides a benchmark for comparison with experimental measurements of the 
forces on moving ions, as well as with more detailed models for the QP-ion interaction.
Model A in Table \ref{table_model} is the hard-sphere potential with radius for the electron bubble
in \He\ at $P=0\,\,\mbox{bar}$, i.e. $k_f R = 11.17$, as determined by the normal-state mobility.
The theoretical results for the forces on a moving electron bubble in \He, and the comparison with 
the experimental data reported in Refs.~\cite{ike13,ike13b,ike15} for the mobility of negative ions 
in normal and superfluid \Hea, is given in Ref. \cite{she16} and summarized in Fig. \ref{Fig1}.
Panel (a) shows the longitudinal mobility as a function of temperature, which is in 
perfect agreement with the experimental data over more than two decades for $0.25 \leq T/T_c < 1$.
From the inset, note that the number of angular momentum channels contributing substantially to 
QP-ion scattering is finite and determined by $l_{\mathrm{max}} \leq k_fR$. For $l > k_fR$ the
phase shifts decrease rapidly to zero. Panel (b) shows the tangent of the Hall angle as a 
function of $\Delta(T)/k_BT$ calculated from the ratio of the transverse and longitudinal 
Stokes parameters [Eq.~(\ref{eq-Hall_angle})]. 
The theory based on the hard sphere model (Model A) is in a good agreement with mobility experiments 
for electron bubbles, providing confirmation that the microscopic theory for potential and branch 
conversion scattering of Bogoliubov quasiparticles captures the essential physics and structure of 
the negative ion moving at low velocity in a chiral superfluid.
It is worth noting that the hard-sphere radius $R$ was fixed at the outset by fitting the 
calculated normal-state mobility to the experimentally measured value, and that there are no 
other adjustable parameters in the calculations for the forces on the ion in superfluid \He.
Nevertheless, it is important to test the robustness of the theoretical predictions by considering 
a range of models for the QP-ion potential, as well as possible variations in transport properties 
for ions described by a potential that deviates signficantly from that of a hard sphere.

\vspace*{-3mm}
\subsection{Piece-wise constant potential with intermediate-range attraction}\label{subsec_ESM}
\vspace*{-3mm}

\begin{table}
\caption{Quasiparticle-Ion potentials $U(r)$}\label{table_model}
\begin{tabular}{lll}
	\hline\noalign{\smallskip}
	Label & Potential & Parameters \\
	\noalign{\smallskip}\hline\noalign{\smallskip}
	Model A & hard sphere & $k_fR = 11.17$ \\
	Model B & repulsive core \& attractive well & $U_0 = 100 E_f, U_1 = 10 E_f, k_fR' = 11, R/R' = 0.36$\\
	Model C & random phase shifts 1 & $l_{\mathrm{max}} = 11$\\
	Model D & random phase shifts 2 & $l_{\mathrm{max}} = 11$\\
	Model E & P\"{o}schl-Teller 1 & $U_0 = 1.01E_f, k_fR = 22.15, \alpha = 3\times 10^{-5}, n = 4$\\
	Model F & P\"{o}schl-Teller 2 & $U_0 = 2 E_f, k_fR = 19.28, \alpha = 6\times 10^{-5}, n = 4$\\
	Model G & hyperbolic tangent 1 & $U_0 = 1.01E_f, k_fR = 14.93, b = 12.47, c = 0.246$\\
	Model H & hyperbolic tangent 2 & $U_0 = 2 E_f, k_fR = 14.18, b = 11.92, c = 0.226$\\
	Model I & soft sphere 1 & $U_0 = 1.01E_f, k_fR = 12.48$\\
	Model J & soft sphere 2 & $U_0 = 2 E_f, k_fR = 11.95$\\
	\noalign{\smallskip}\hline
	\end{tabular}
\end{table}

Among the analytically solvable models we consider Model B for the QP-ion potential with finite, 
short-range repulsion and intermediate range attraction defined by the piece-wise constant potential,
\be
U(r) = 
\begin{cases}
U_0, 	& r < R		\,,\\
-U_1, 	& R < r < R'	\,,\\
0, 	& r > R'	\,.
\end{cases}
\ee
Using Eq.~(\ref{phase_shifts_def}), we find the following expression for the phase shifts,
\ber
&&
\tan\delta_l
=
\frac{(l-\zeta_l)j_l(k_fR')-k_fR'j_{l+1}(k_fR')}{(l-\zeta_l)n_l(k_fR')-k_fR'n_{l+1 }(k_fR')}
\,,\quad
\zeta_l = x'\frac{a_l}{b_l}.
\label{sswaw}
\\
a_l &=& l\left[n_{l+1}(x)j_l(x') - n_l(x')j_{l+1}(x)\right]
+ x'\left[n_{l+1}(x')j_{l+1}(x) - n_{l+1}(x)j_{l+1}(x')\right]
\notag\\
&+& \frac{lp_l}{x'}\left[n_l(x)j_l(x') - n_l(x')j_l(x)\right]
+ p_l\left[n_{l+1}(x')j_l(x) - n_l(x)j_{l+1}(x')\right],\\
b_l &=& x'\left[n_{l+1}(x)j_l(x') - n_l(x')j_{l+1}(x)\right]
+ p_l\left[n_l(x)j_l(x') - n_l(x')j_l(x)\right],
\label{eq-deltas-betas}
\eer
with $p_l = z'i_{l+1}(z)/i_l(z)$, $x = \beta_1k_fR$, $x' = \beta_1k_fR'$, 
$z = \beta_0k_fR$, $z' = \beta_0k_fR'$, $\beta_0 = \sqrt{(U_0-E_f)/E_f}$ and 
$\beta_1 = \sqrt{(U_1+E_f)/E_f}$, where $i_l(x)$ is the modified spherical Bessel function of the first 
kind. 
Note that a purely repulsive soft-core potential is obtained from Eq.~(\ref{sswaw}) by 
setting $R = R'$ and $U_1 = 0$.

\begin{figure}
\centering
\includegraphics[width=1.0\textwidth,keepaspectratio]{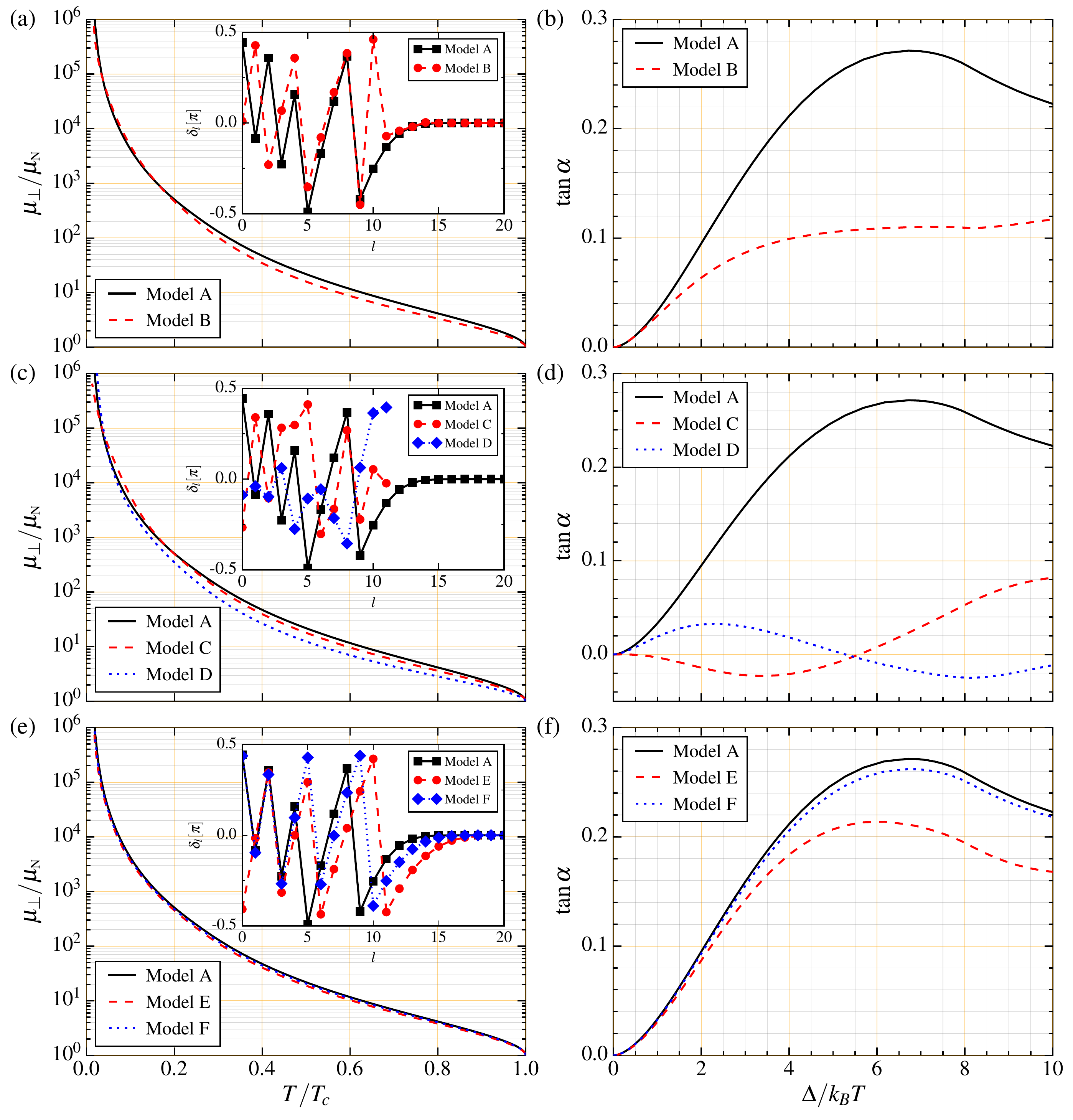}
\caption{Comparison of numerical results for longitudinal mobility ($\mu_{\perp}/\mu_{\text{N}}$) 
and the Hall ratio ($\tan\alpha=\eta_{\text{AH}}/\eta_{\perp}$) obtained with Models B-F listed
in Table \ref{table_model}, in comparison with the hard-sphere model (Model A). 
(Color figure online.)}
\label{Fig2}
\end{figure}

\vspace*{-3mm}
\paragraph{Model B vs Model A.}
In Figs.~\ref{Fig2} (a)-(b) we compare calculations based on Model A with those based on Model B 
(parameters are listed in Table \ref{table_model}).
Model A is the hard sphere that agrees very well with experiments on the electron bubble in \Hea.
Model B corresponds to strong short-range repulsion, and intermediate range attraction. The
latter allows for a shallow bound state, and therefore a scattering resonance, in one or more high 
angular momentum channels.
In the case of Model B one can see that there is an \emph{extra} scattering resonance 
in channel $l = 10$ shown in the inset of Fig.~\ref{Fig2} (a).
Figure \ref{Fig2} (b) shows that the resonance leads to small deviations of the drag force compared to 
that for the hard-sphere model (and experiment), but a drastic reduction in the Hall ratio.
The basic conclusion is that the electron bubble is well described by strong, short-range repulsion
with no intermediate range attraction.
A variant of Model B may be relevant to positive ions, since postive ions attract \He\ atoms 
producing a more complex ionic structure.

\vspace*{-3mm}
\subsection{Other QP-ion Potentials and Scattering Models}\label{subsec_OM}
\vspace*{-3mm}

Apart from the two exactly solvable models discussed above, we considered several 
repulsive potentials, as well as a constrained random-phase-shift model 
(Models C and D in Table \ref{table_model}). 

\vspace*{-3mm}
\paragraph{Models C and D.}
The pattern of phase shifts versus angular momentum for Model A shown in the inset of Fig.~\ref{Fig1}
is a fingerprint of the hard sphere QP-ion potential, \emph{with} $k_f R = 11.17$ fixed by the 
normal-state ion mobility. This constraint fixes the number of relevant scattering channels.
Here we consider the sensitivity of forces on the ion to the specific pattern of phase shifts, 
while keeping the number of scattering channels fixed at $l_{\text{max}}=11$ \emph{and} enforcing 
the constraint on the transport cross section provided by the normal-state mobility. 
Models C and D are two different realizations of the \emph{random-phase-shift model} with 
$l_{\text{max}}=11$, constrained to fit the normal-state ion mobility, $\mu_{\text{N}}$. 
The procedure is to use a random number generator to calculate $\{\delta_l| l = 1,\dots,l_{\mathrm{max}}\}$, 
then adjust the phase shift in channel $l=0$ to satisfy the constraint on the transport cross section. 
Models C and D differ by the seed used to generate the 
phase shifts. Note that not every realization of random phase shifts for $l = 1,\dots,l_{\text{max}}$ 
allows a fit to experiment by varying the remaining phase shift $\delta_0$.
As Fig.~\ref{Fig2} (d) shows the random-phase-shift model fails dramatically to account for 
the anomalous Hall angle for the electron bubble in \Hea, and thus the magnitude and temperature
dependence of the transverse force on the electron bubble, even though the longitudinal mobility 
shown in Fig.~\ref{Fig2} (c) is relatively close to that of Model A, and therefore to the measured
longitudinal force.
This basic feature is characteristic of the comparison between theory and mobility measurements 
for electron bubbles in \Hea; the longitudinal mobility is relatively insensitive to the QP-ion 
potential provided the model accounts for the experimental normal-state transport cross section.
In contrast, the transverse force is sensitive to the pattern of phase shifts as a function of 
the angular momentum channel, as well as the transport cross section. 
  
\vspace*{-3mm}
\paragraph{Models E and F.}
Additional motivation for considering a range of models for the QP-ion potential is to see if a 
refinement to Model A can remove the small deviations between theory and experiment evident in 
Fig.~\ref{Fig1} for the the Hall ratio at temperatures near $T_c$, i.e. for $\Delta(T)/k_B T\rightarrow 0$. 
Thus, we consider P\"{o}schl-Teller potentials of the form $U(x) = U_0/\cosh^2[\alpha x^n]$, 
where $x=k_f r$, as well as the hyperbolic tangent model defined by $U(x)=U_0[1-\tanh[(x-b)/c]]$.
In all cases the parameters of the potential are adjusted to fit the normal-state transport cross 
section to account for the measured normal-state mobility, 
$\mu_{\text{N}}^{\mathrm{exp}}=1.7\times10^{-6}\,\mathrm{m^2/Vs}$ \cite{ike13}.

In Figs.~\ref{Fig2} (e)-(f) we show numerical results for the P\"{o}schl-Teller model with 
two different sets of parameters as listed in Table~\ref{table_model}. This 
is a three-parameter model describing a smoothly decaying repulsive potential. As was the case
for other models the transverse component of the Stokes tensor, $\eta_{\text{AH}}$, and 
thus the Hall angle, is more sensitive to the structure of the potential [Fig.~\ref{Fig2}(f)], 
than is the longitudinal mobility $\mu_{\perp}$. 
The phase shifts, particularly those of Model F, are very close to those of Model A [inset 
of Fig.~\ref{Fig2} (e)]. The results for Model F are also much closer to Model A, and to experiment, 
than those of Model E, which is a softer and longer range potential.
The general trend is that the numerical results obtained with the P\"{o}schl-Teller 
potential are almost indistinguishable from Model A in the limit that the strength of the potential
is sufficiently repulsive, i.e. $U_0 \gtrsim 6E_f$ (not shown).

\vspace*{-3mm}
\paragraph{Models G, H, I and J.}
Lastly, we consider the hyperbolic tangent and the soft-sphere models, each with 
two different sets of parameters, indicated in Table~\ref{table_model} as Models G and H
and Models I and J, respectively. 
The numerical results, which are not shown, for these models are barely distinguishable from 
those of Model A, and in contrast to the P\"{o}schl-Teller model, the results for 
Models G-J are practically insensitive to variations of the magnitude of the potential $U_0$, 
provided $U_0>E_f$.
Finally, we note that the small deviations between theory and experiment for the anomalous Hall 
ratio persist, suggesting that there may be an additional scattering mechanism not captured by 
a repulsive, short-range, spin-independent, isotropic QP-ion potential. 
 
\vspace*{-3mm}
\section{Conclusions}\label{sec_Concl}

We considered a number of models for the effective potential describing the interaction
between normal-state quasiparticles and ions embedded in \He. This potential determines the 
normal-state scattering phase shifts which are the input parameters to the theory for Bogoliubov
quasiparticles scattering off electron bubbles moving in the chiral A phase of superfluid \He. 
We show that the scattering theory developed in Ref. \cite{she16}, with a strongly repulsive,
short-range QP-ion potential - specifically the hard sphere model - is in very good agreement 
with experimental measurements of the longitudinal mobility and anomalous Hall current for 
electron bubbles moving in \Hea\ \cite{ike13,ike13b,ike15}. 
We also show that softer, longer range potentials, as well as potentials with intermediate 
range attraction, fail to account for the magnitude and temperature depedence of the Hall angle.

\vspace*{-3mm}
\begin{acknowledgements}
The research of OS and JAS was supported by the National Science Foundation (Grant 
DMR-1508730). We thank Hiroki Ikegami, Kimitoshi Kono and Yasumasa Tsutsumi for 
discussions on their mobility experiments and interpretaions. 
\end{acknowledgements}

%

\end{document}